\documentclass[conference,10pt]{IEEEtran}
\usepackage{epsfig,epsf,rotating,setspace,latexsym,amsmath,amssymb,amsfonts,bm,theorem,subfigure,epstopdf,cite,authblk,bbm,comment,enumitem}
\usepackage{algorithm}
\usepackage[noend]{algpseudocode}
\usepackage{color}
\usepackage{mathtools}
\usepackage{soul}
\usepackage{multicol}

\algrenewcommand\algorithmicforall{\textbf{foreach}}
\algrenewcommand\algorithmicindent{.8em}

\newlist{notes}{enumerate}{1}
\setlist[notes]{label=Note: ,leftmargin=*}

\newtheorem{remark}{Remark}

\newsavebox{\shortpagebox}

\makeatletter
\newcommand{\shortpage}[1]
{\par
  \setbox\shortpagebox=\vbox{\strut #1\par}%
  \afterpage{\onecolumn
    \begin{multicols}{2}
    \unvbox\AP@partial
    \end{multicols}}%
  \unvbox\shortpagebox
\par}
\makeatother

\IEEEoverridecommandlockouts
\allowdisplaybreaks

\begin{document}

\title{Mobility in Age-Based Gossip Networks}
 
\author{Arunabh Srivastava \qquad Sennur Ulukus\\
        \normalsize Department of Electrical and Computer Engineering\\
        \normalsize University of Maryland, College Park, MD 20742\\
        \normalsize  \emph{arunabh@umd.edu} \qquad \emph{ulukus@umd.edu}}

\maketitle

\begin{abstract}
    We consider a gossiping network where a source forwards updates to a set of $n$ gossiping nodes that are placed in an arbitrary graph structure and gossip with their neighbors. In this paper, we analyze how mobility of nodes affects the freshness of nodes in the gossiping network. To model mobility, we let nodes randomly exchange positions with other nodes in the network. The position of the node determines how the node interacts with the rest of the network. In order to quantify information freshness, we use the version age of information metric. We use the stochastic hybrid system (SHS) framework to derive recursive equations to find the version age for a set of positions in the network in terms of the version ages of sets of positions that are one larger or of the same size. We use these recursive equations to find an upper bound for the average version age of a node in two example networks. We show that mobility can decrease the version age of nodes in a disconnected network from linear scaling in $n$ to at most square root scaling and even to constant scaling in some cases. We perform numerical simulations to analyze how mobility affects the version age of different positions in the network and also show that the upper bounds obtained for the example networks are tight.
\end{abstract}

\section{Introduction}
\begin{figure}[t]
    \centering
    \includegraphics[scale=0.5]{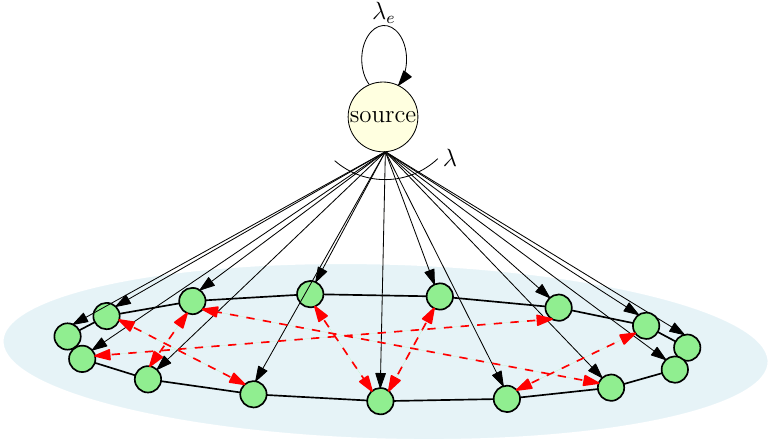}
    \caption{A gossiping network where each node can gossip with its neighbors, denoted by black lines, and exchange positions with some nodes in the network, denoted by red arrows.}
    \label{fig:system model}
\end{figure}

The field of wireless communication has seen transformative changes in recent years. The rollout of the 5G networks for the general public has enhanced the connectivity and productivity of the human population. With these technologies helping us with full force, we look for ways to further improve communication technology for new use cases. In this paper, we focus on a specific use case of having mobility in dense networks of users; see Fig.~\ref{fig:system model}. The use of swarm drones and self-driving cars in performing time-critical tasks is becoming increasingly viable. Hence, we analyze how such mobile networks perform from the perspective of freshness of information.

Using latency and throughput as metrics to analyze networks is not sufficient to quantify freshness of information of nodes in the network\cite{popovski2022perspective}. New metrics have been proposed to quantify freshness of information, such as age of information \cite{kaul2012real, sun2019age, yatesJSACsurvey}. Several extended metrics have also been introduced based on real-life inspired applications, including age of incorrect information\cite{maatouk20AOII}, age of synchronization \cite{zhong18AoSync}, binary freshness metric \cite{cho3BinaryFreshness}, and version age of information \cite{yates21gossip, Abolhassani21version, melih2020infocom}.

In this paper, we use the version age of information metric to quantify freshness of information. The version age of a node in the gossiping network is the number of versions behind a node is when compared to the source node generating or observing a random process and forwarding the latest updates to the network. The SHS framework was used in \cite{yates21gossip} for the first time to derive recursive equations to find version age of sets of nodes in terms of one larger sets. \cite{yates21gossip} also found that the average version age scaling of a single node in a fully-connected network is logarithmic. \cite{buyukates22ClusterGossip, srivastava2023grid, srivastava2023generalizedrings, maranzatto24} analyze different graph structures and find average version age scaling. \cite{kaswan2023versionagesurvey} reviews recent works in the area of gossiping for freshness.

Many works have analyzed networks with mobility using age of information as a metric. \cite{chaintreau2009age} considers a gossiping network consisting of multiple classes which nodes can move in and out of. The paper uses the spatial mean field regime from statistical mechanics to find a differential equation that gives the joint distribution of age of information and time. This work considers that each node in the network can access a base station with a fixed rate irrespective of the number of nodes in the network. This is different from the model we consider in this paper, where the rate of receiving information from the source node decreases as the number of nodes in the network increases. \cite{tripathi2019age} considers the problem of minimizing age of information and UAV trajectory design for both information gathering and information dissemination problems. It considers both networks with mobile nodes and networks with stationary nodes and mobile unmanned aerial vehicles (UAVs). The mobility models considered are random walk mobility and i.i.d.~mobility. \cite{wu21uav} considers the problem of trajectory planning for multiple relay UAVs while minimizing AoI and uses reinforcement learning. \cite{eslam23uav} considers a similar problem and includes energy consumption minimization along with trajectory planning and age minimization.

In this work, we consider a general gossiping network that receives updates from a source that has the latest version of the updates. To model mobility, we let nodes in the network exchange positions with other nodes depending on their position in the network. We find recursive equations that give the version age of a set of positions in the gossiping network in terms of sets that are one larger, i.e., have one more position than the set whose version age we find, and sets with the same number of positions. We analyze how mobility affects the version age of positions in the network by considering two different toy examples. Then, we analyze two constructed gossiping networks and use the recursive equations we obtained to find a tight upper bound for the average version age of a single node in each network. Finally, we carry out numerical simulations to verify and analyze our results.

\section{System Model} \label{section 2}
We have a wireless network consisting of a source node and a gossip network where a source node generates or observes updates as a rate $\lambda_e$ Poisson process independent of all other processes in the network. An illustration is shown in Fig.~\ref{fig:system model}. The set of nodes in the gossiping network is denoted by $\mathcal{N}$ where $|\mathcal{N}| = n$. The nodes are placed in arbitrary fixed positions in the network at the start of the process. We name these positions as $[1,2,\ldots,n]$ serially starting at an arbitrary position. We name the nodes in each position at the start of the process with the same label as the position. The source node also shares its updates with nodes in the gossiping network as a combined rate $\lambda$ Poisson process. The source sends updates to position $j$ as a rate $\lambda_{0j}$ Poisson process, independent of all processes in the network. Thus, $\sum_{j \in \mathcal{N}} \lambda_{0j} = \lambda$.

Each node in the gossiping network can communicate with its neighbors as a Poisson process independent of all other processes in the network. The set of neighbors each node can communicate with depends on the position it occupies in the network. If a node in position $i$ sends an update to a node in position $j$, then the rate of information flow, i.e., the rate of the Poisson process of updates sent from position $i$ to $j$ is $\lambda_{ij}$.

Next, we define the version age of information metric, which we use to quantify the freshness of version updates in the gossiping network. Suppose $N_0(t)$ is the counting process associated with the version updates at the source node. Then, $N_0(t)$ increases by $1$ each time the source gets new version updates. Similarly, we define the counting process $N_i(t)$ to represent the version updates of position $i$, thus maintaining the latest update present at $i$. The version age of position $i$ at time $t$, denoted by $X_i(t)$, is defined as the number of versions behind the update at the node at position $i$ is when compared to the source node and is represented mathematically as $X_i(t) = N_0(t) - N_i(t)$.  The version age of a set $S$ of fixed positions in the network is defined as $X_S(t) = \min_{j \in S}X_j(t)$. We  define the expected value and limit as $v_S(t) = \mathbb{E}[X_S(t)]$ and $v_S = \lim_{t \rightarrow \infty} v_S(t)$, respectively.

We say that position $i$ is a neighboring position of set $S$ if $\lambda_{ij} > 0$ for some $j \in S$, and define the set of neighboring positions of $S$ as $N(S)$. We define the rate of information flow from node in position $i$ into set $S$ as $\lambda_i(S) = \sum_{j \in S}\lambda_{ij}$ if $i \notin S$ and $\lambda_i(S) = 0$ if $i \in S$. Similarly, we define the rate of information flow from the source to set $S$ as $\lambda_0(S)$.

\begin{figure}
    \centering
    \includegraphics[scale = 0.5]{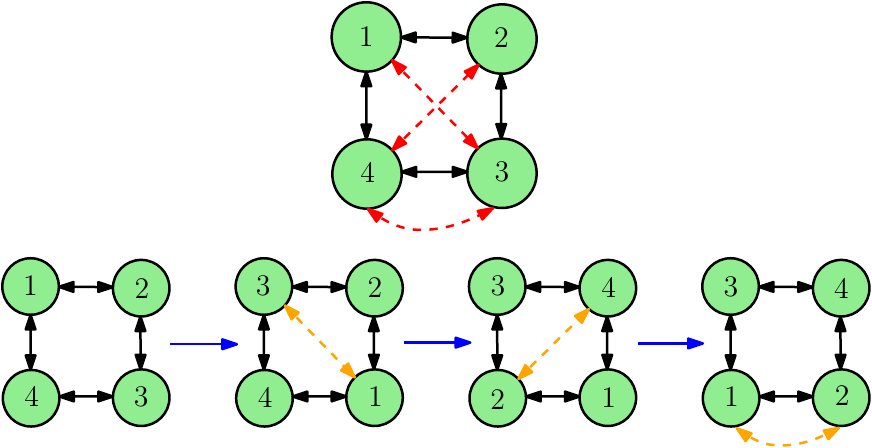}
    \caption{An illustration of the mobility model in the gossiping network. The positions where node exchanges are possible are denoted by red arrows in the network on the top. As shown by the yellow arrows in the process, nodes $1$ and $3$ exchange positions, followed by nodes $2$ and $4$, and finally nodes $1$ and $2$ exchange positions.}
    \label{fig:mobility understanding}
\end{figure}

Finally, we define the mobility processes in the gossiping network. A node in position $i$ in the network can exchange positions with nodes in a subset $M_i$ of the other positions in the network. The set $M_i$ may be empty. We assume that all node exchange processes are independent of each other and every other process in the network. For any position $i$ in the network and any position $j \in M_i$, the exchange process is defined as a rate $\lambda_{ij}^m$ Poisson process, with superscript $m$ signifying \emph{mobility}. Hence, each node exchange depends on the positions, but not on the nodes present in this position. The node exchange process is illustrated in Fig.~\ref{fig:mobility understanding}. We say that a gossip network has full mobility if a node in any position can exchange positions with any other node in the network.

\section{SHS Characterization of the Network}
In this section, we use the SHS characterization following \cite{yates21gossip} and \cite{hespanha_SHS} and find recursive equations to find $v_S$ for any general set of positions in the network. We notice that the rates of all Poisson processes in the network are constant with respect to time and that all processes are memoryless. Hence, we need only one discrete state to represent the network. We define the discrete state as $\mathcal{Q} = \{0\}$ and do not use notation to represent it as a function input later. We define the continuous state of the system to be a vector of version ages of positions $1$ through $n$, i.e. $\mathbf{X}(t) = \left[X_1(t), X_2(t), \ldots, X_n(t) \right]$. We see that all changes in the continuous state are piecewise constant and only change when a transition happens. Hence, we can write the stochastic differential equation as $\Dot{\mathbf{X}}(t) = 0$.

Next, we define the transition/reset maps $\mathcal{L}$ of the SHS. We can define a transition in the network uniquely using three variables $(i,j,k)$ where $i$ is the position that initiates the action, $j$ is the second position that participates, and $k$ tells us whether the transition is related to gossiping or a node exchange.  Let $(i,j,
\text{info})$ denote the transitions which lead to information exchanges, corresponding to the gossiping processes where node $i$ updates node $j$, and $(i,j,\text{node})$ denote transitions that lead to node exchanges. The source related transitions are different from the regular gossip transitions. We have two types of source related transitions. The first type is when the source updates itself. We denote this as $(0,0,\text{info})$. The second type of transition happens when the source sends an update to one of the positions. The transition that occurs when the source sends an update to position $j$ is denoted by $(0,j,\text{info})$. Hence, the entire set of transitions can be represented by the set
\begin{align}
    \mathcal{L} = & \{(0,0,\text{info})\} \cup \{(0,j,\text{info}): j \in \mathcal{N}\} \notag\\
    &\cup \{(i,j,\text{info}): i,j \in \mathcal{N}\} \notag \\
    &\cup \{(i,j,\text{node}): i,j \in \mathcal{N}, i<j\}.
\end{align}
Here, we note that we chose the transitions denoting node exchanges such that we can uniquely identify each transition. In general, node exchanges are not initiated by nodes, and hence the transition is symmetric in both nodes participating in the exchange. Let the continuous state after a transition be $\phi_{i,j,k}(\mathbf{X(t)},t) = \left[X_1'(t), X_2'(t), \ldots, X_n'(t)\right]$. Then, we have
\begin{align}
    \!\!\!X_l'(t) = \begin{cases}
            X_l(t) + 1, & i = 0, j = 0, k = \text{info}\\
            0, & i = 0, j = l, k = \text{info} \\
            \min(X_l(t),X_i(t)), & i\in \mathcal{N}, j = l, k = \text{info}\\
            X_i(t), & i \in \mathcal{N}, l \in M_i, k = \text{node}\\
            X_l(t), & \text{otherwise}. \!\!
    \end{cases}
\end{align}
The transition rates corresponding to the transition maps are 
\begin{align}
    \lambda_{i,j,k}(\mathbf{X}(t),t) = \begin{cases}
            \lambda_e, & i = 0, j = 0, k = \text{info}\\
            \lambda_{0l}, & i = 0, j\in \mathcal{N}, k = \text{info} \\
            \lambda_{ij}, & i,j \in \mathcal{N}, k = \text{info}\\
            \lambda_{ij}^m, & i \in \mathcal{N}, j \in M_i, k = \text{node}.
    \end{cases}
\end{align}

Now, we choose a test function $\psi_S: \mathbb{R} \times [0,\infty) \rightarrow \mathbb{R}$ as $\psi_S(\mathbf{X}(t),t) = X_S(t)$, where $S$ is a set of positions in the network that can  never include the position of the source. Now, we write the extended generator equation,
\begin{align}\label{eq: extended generator equation}
    \!\!\!\!(L\psi_S)(\mathbf{X}(t)) = \sum_{i,j,k}(\psi_S(\phi_{i,j,k}(\mathbf{X}(t)))-\psi_S(\mathbf{X}(t)))\lambda_{i,j,k}. \!
\end{align}
Since $X_S(t)$ is piece-wise constant, we get $\frac{\partial X_S(t)}{\partial t} = 0$ and arrive at the above equation.

The values of $\left[ \psi_S(\phi_{i,j,k}(\mathbf{X})) \right]$ are: $0$ if $i = 0, j \in S, k = \text{info}$, $X_S(t) + 1$ if $i = 0, j = 0, k = \text{info}$, $X_{S \cup \{i\}}(t)$ if $i \in N(S), j \in S, k = \text{info}$, $X_{S \cup \{i\} \backslash \{l\}}(t)$ if $i \in M_l, j = l, k = \text{node}$ and $X_S(t)$ otherwise. 

We substitute this in \eqref{eq: extended generator equation}, take expectation on both sides, choose $t \rightarrow \infty$, and use Dynkin's Formula to obtain
\begin{align}
    0 = &\lambda_e(v_S + 1 - v_S) + \sum_{i \in S}\lambda_{0j}(0 - v_S) \notag \\
    & + \sum_{i \in N(S)} \lambda_i(S) (v_{S \cup \{i\}} - v_S) \notag \\
    & + \sum_{i \in S} \sum_{j \in M_i \backslash S} \lambda_{ij}^m(v_{S \cup \{j\} \backslash \{i\}} - v_S) \label{eq: recursion in line}.
\end{align}
We collect the terms of $v_S$ on one side, rewrite $v_{S \cup \{j\} \backslash \{i\}}$ as $v_{S(i,j)}$, $M_i \backslash S = M_i^S$, and rewrite the equation to get
\begin{align}
    v_S = \frac{\lambda_e + \sum_{i \in N(S)} \lambda_i(S)v_{S \cup \{i\}} + \sum_{i \in S} \sum_{j \in M_i^S} \lambda_{ij}^m v_{S(i,j)}}{\lambda_0(S) + \sum_{i \in N(S)} \lambda_i(S) + \sum_{i \in S} \sum_{j \in M_i^S} \lambda_{ij}^m} \label{eq: recursive equation}
\end{align}
This representation using positions gives rise to simultaneous recursive equations of sets of positions with the same size, i.e., same number of positions in the set, in terms of sets of positions that are one larger. In contrast, when there is no mobility, the version age of a set of nodes $S$ depends only on the version age of sets that can be constructed by adding a neighbor of $S$ to it\cite{yates21gossip}. After this point, whenever we mention version age, we mean the limiting average version age.

\begin{remark}
    The limiting average version age of a position in the network is the same as the limiting average version age of a node in the network. Hence, we can use the recursive equations in \eqref{eq: recursive equation} to find the average version age of a node.
\end{remark}

\section{A Toy Example}\label{sec: toy example}
\begin{figure}
    \centering
    \includegraphics[scale = 0.4]{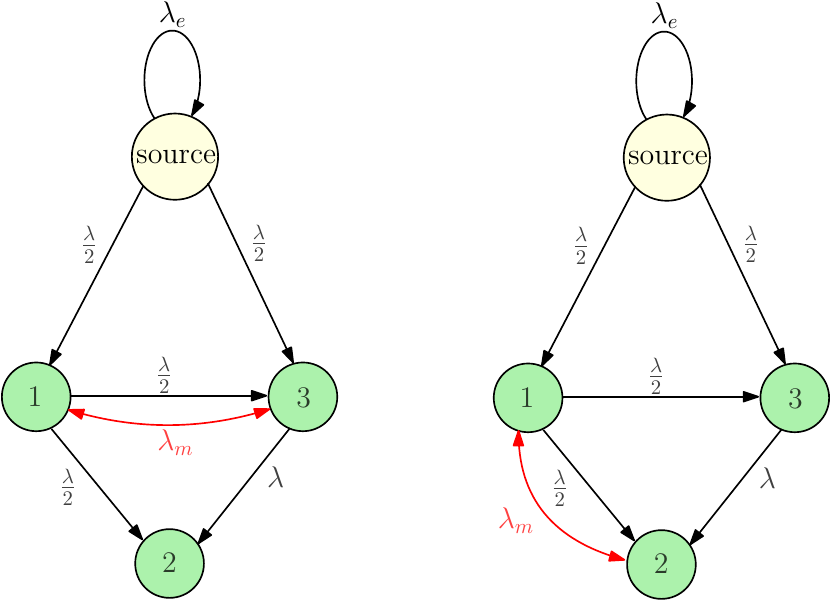}
    \caption{Toy example in \cite{yates21gossip}, with mobility. Left: nodes in positions $1$~and~$3$ can exchange positions. Right: nodes in positions $1$~and~$2$ can exchange positions.}
    \label{fig: toy example mobility}
    \vspace*{-0.4cm}
\end{figure}

We consider the toy example solved in \cite{yates21gossip} and apply mobility to it in two ways. This is shown in Fig.~\ref{fig: toy example mobility}. We want to see how the version ages of positions are affected by mobility. In order to simplify calculations, we assume that the source sends updates as a combined rate $\lambda$, and the nodes in positions $1$ and $3$ send updates as a combined rate $\lambda$ process. Moreover, the rate of node exchange between positions is $\lambda_m$.

We recall from \cite{yates21gossip} that the version age of nodes/positions in the network if there was no mobility are
\begin{align}\label{eq: no mobility toy}
    v_1 = 2\frac{\lambda_e}{\lambda}, \qquad v_2 = 2\frac{\lambda_e}{\lambda}, \qquad v_3 = \frac{3}{2} \frac{\lambda_e}{\lambda}.
\end{align}

First, we solve the equations for the network where node exchange can happen between positions $1$ and $3$. Using \eqref{eq: recursive equation}, we write equations going from $\{1,2,3\}$ to the smallest sets, 
\begin{align}
    v_{\{1,2,3\}} =& \frac{\lambda_e}{\lambda},\\
    v_{\{1,2\}} =& \frac{\lambda_e + \lambda v_{\{1,2,3\}} + \lambda_m v_{\{2,3\}}}{\frac{\lambda}{2} + \lambda + \lambda_m}\\
    =& \frac{2\lambda_e + \lambda_m v_{\{2,3\}}}{\frac{3}{2}\lambda + \lambda_m},\\
    v_{\{2,3\}} =& \frac{\lambda_e + \lambda v_{\{1,2,3\}} + \lambda_m v_{\{1,2\}}}{\frac{\lambda}{2} + \lambda + \lambda_m}\\
    =& \frac{2\lambda_e + \lambda_m v_{\{1,2\}}}{\frac{3}{2}\lambda + \lambda_m},\\
    v_{\{1,3\}} =& \frac{\lambda_e}{\lambda}.
\end{align}
Using these, we find the version age of the three positions in the network as follows,
\begin{align}
    v_1 =& \frac{\lambda_e}{\lambda}\frac{\lambda+\frac{5}{2}\lambda_m}{\frac{\lambda}{2}+\frac{3}{2}\lambda_m},\\
    v_3 =& \frac{\lambda_e}{\lambda}\frac{\frac{3}{4}\lambda+\frac{5}{2}\lambda_m}{\frac{\lambda}{2}+\frac{3}{2}\lambda_m},\\
    v_2 =& 2\frac{\lambda_e}{\lambda}.
\end{align}

Similarly, we solve the equations for the network where node exchange can happen between positions $1$ and $2$. Using \eqref{eq: recursive equation}, we write equations going from the largest to smallest sets, 
\begin{align}
    v_{\{1,2\}} =&  \frac{\lambda_e + \lambda v_{\{1,2,3\}}}{\frac{\lambda}{2} + \lambda} \\
    =& \frac{4\lambda_e}{3\lambda},\\
    v_{\{2,3\}} =& \frac{\lambda_e + \lambda v_{\{1,2,3\}} + \lambda_mv_{\{1,3\}}}{\frac{\lambda}{2} + \lambda + \lambda_m}\\
    =& \frac{2\lambda_e + \lambda_mv_{\{1,3\}}}{\frac{3\lambda}{2} + \lambda_m},\\
    v_{\{1,3\}} =& \frac{\lambda_e + \lambda_mv_{\{2,3\}}}{\lambda + \lambda_m}.
\end{align}
Solving these simultaneous equations together, we get
\begin{align}
    v_{\{2,3\}} =& \frac{\lambda_e}{\lambda}\frac{2\lambda+3\lambda_m}{\frac{3}{2}\lambda+\frac{5}{2}\lambda_m},\\
    v_{\{1,3\}} =& \frac{\lambda_e}{\lambda}\frac{\frac{3}{2}\lambda+3\lambda_m}{\frac{3}{2}\lambda+\frac{5}{2}\lambda_m},
\end{align}
and therefore,
\begin{align}
    v_2 =& \frac{\lambda_e + \frac{\lambda}{2}v_{\{1,2\}} + \lambda v_{\{2,3\}} + \lambda_m v_1}{3\frac{\lambda}{2} + \lambda_m}\\
    =& \frac{\lambda_e}{\lambda}\frac{\frac{\lambda}{2}+\lambda_m}{\frac{3}{4}\lambda+2\lambda_m}\left(\frac{5}{3}\!+\!\frac{2\lambda+3\lambda_m}{\frac{3}{2}\lambda+\frac{5}{2}\lambda_m}\!+\!\frac{\lambda_m}{\frac{\lambda}{2}+\lambda_m}\right),\\
    v_1 =& \frac{\lambda_e + \lambda_mv_2}{\frac{\lambda}{2} + \lambda_m}\\
    =& \frac{\lambda_e}{\frac{\lambda}{2} + \lambda_m}+\frac{\lambda_e}{\lambda}\frac{\lambda_m}{\frac{3}{4}\lambda+2\lambda_m}\left(\frac{5}{3}\!+\!\frac{2\lambda+3\lambda_m}{\frac{3}{2}\lambda+\frac{5}{2}\lambda_m}\!+\!\frac{\lambda_m}{\frac{\lambda}{2}+\lambda_m}\right),\\
    v_3 =& \frac{\lambda_e + \frac{\lambda}{2}v_{\{1,3\}}}{\lambda} = \frac{\lambda_e}{\lambda}\frac{\frac{9}{2}\lambda+8\lambda_m}{3\lambda+5\lambda_m}.
\end{align}

We first observe that we get the results in \eqref{eq: no mobility toy} if we substitute $\lambda_m = 0$ for both toy examples.

Next, we observe that, if we choose $\lambda_m \rightarrow \infty$, we obtain, for the case where $1$ and $3$ can exchange positions,
\begin{align}\label{eq: 34}
v_1 = \frac{5}{3}\frac{\lambda_e}{\lambda}, \qquad v_2 = 2\frac{\lambda_e}{\lambda}, \qquad v_3 = \frac{5}{3}\frac{\lambda_e}{\lambda},
\end{align}
and for the case where $1$ and $2$ can exchange positions, 
\begin{align}\label{eq: 35}
v_1 = \frac{29}{15}\frac{\lambda_e}{\lambda}, \qquad v_2 = \frac{29}{15}\frac{\lambda_e}{\lambda}, \qquad v_3 = \frac{8}{5}\lambda. 
\end{align}
The first thing we notice from observing the increase in version age $v_3$ in \eqref{eq: 34} and \eqref{eq: 35} when compared to the no mobility case in \eqref{eq: no mobility toy} is that mobility does not improve the version age of every position in the network. Further, when there is no mobility, the average version age in \eqref{eq: no mobility toy} is $\frac{11}{6}\frac{\lambda_e}{\lambda}=1.833\frac{\lambda_e}{\lambda}$. When there is mobility where $1$ and $3$ can exchange positions, the average version age in \eqref{eq: 34} is $\frac{16}{9}\frac{\lambda_e}{\lambda}=1.778\frac{\lambda_e}{\lambda}$; and when $1$ and $2$ can exchange positions, the average version age in \eqref{eq: 35} is $\frac{82}{45}\frac{\lambda_e}{\lambda}=1.822\frac{\lambda_e}{\lambda}$. Hence, we observe that the average version age has decreased in both cases of mobility when compared to the network without mobility in this example. We will analyze this further in Section~\ref{sec: numerical simulations}.

\section{Fully Connected Network with a Single Disconnected Node} \label{sec app}
\begin{figure}[t]
    \centering
    \includegraphics[scale = 0.4]{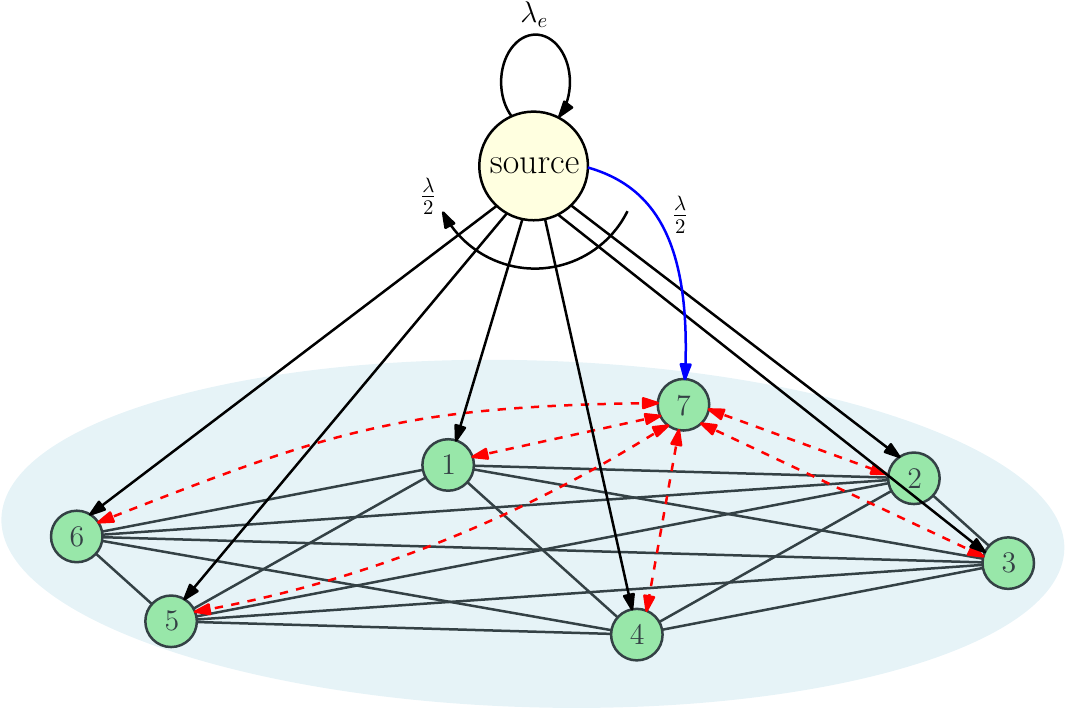}
    \caption{A gossiping network where $\mathcal{N}_{FC}$ is a fully connected network of $6$ nodes $\{1, 2, 3, 4, 5, 6\}$ and $\mathcal{N}_{s}$, represented by position $7$, is a single node. Both parts receive updates as a rate $\frac{\lambda}{2}$ Poisson process from the source. Further, every node in $\mathcal{N}_{FC}$ can exchange positions with the node in $\mathcal{N}_{s}$, as shown by the red arrows.}
    \label{fig: application figure}
\end{figure}

In this section, we apply the recursive equations found in \eqref{eq: recursive equation} to find an upper bound for the version age of the gossiping network described in Fig.~\ref{fig: application figure}. The gossiping nodes are divided into two parts, one part contains a fully connected network of $n-1$ nodes, which we call $\mathcal{N}_{FC}$, and the other part contains one node, which we call $\mathcal{N}_{s}$. We label the positions in $\mathcal{N}_{FC}$ as $\{1,\ldots,n-1\}$ in order and the single node in $\mathcal{N}_{s}$ as $n$. Both parts are served with the same rate from the source, i.e., the source sends updates to each part as a rate $\frac{\lambda}{2}$ Poisson process. Further, the position of $\mathcal{N}_{s}$ can access and be accessed by all nodes in $\mathcal{N}_{FC}$ with rate $\lambda_m = \lambda$ for each node exchange. Hence, $M_j = \{n\}$, $j \in \{1,2,\ldots,n-1\}$, and $M_n = \{1,2,\ldots,n-1\}$. Moreover, $\lambda_{in}^m = \lambda_{ni}^m = \lambda$, $i \in \{1,2,\ldots,n-1\}$. We find an upper bound for the average version age of a single node in the network in this section. We compare it with the average version age in the same network when there is no mobility in Section~\ref{sec: numerical simulations}. 

Due to the symmetry of the network, each position in $\mathcal{N}_{FC}$ of the network has the same version age. Moreover, each set of the same size in $\mathcal{N}_{FC}$ has the same version age. It is easy to see that the version age of the single position in $\mathcal{N}_{s}$ is lower than the positions in $\mathcal{N}_{FC}$, since it receives updates directly from the source much more frequently. Next, due to the symmetry in the network, a set consisting of the single position in $\mathcal{N}_{s}$ and some positions in $\mathcal{N}_{FC}$ will have the same average version age as long as the same number of positions is chosen in $\mathcal{N}_{FC}$. Hence, if we are given that the size of the set of positions is $a$, then there are two types of sets possible. The first type of set is the one which includes $a$ positions from $\mathcal{N}_{FC}$. The second type of set includes $a-1$ positions in $\mathcal{N}_{FC}$ and the single position in $\mathcal{N}_{s}$. Let the representation of the set that has $\alpha$ positions in $\mathcal{N}_{FC}$ and $\beta$ positions in $\mathcal{N}_{s}$ be $(\alpha,\beta)$. Then, we can write down the recursive equations for the two types of sets that can be created with $a$ positions as
\begin{align}
    \!\!\!v_{(a-1,1)} = &\frac{\lambda_e + \frac{(a-1)(n-a-2)\lambda}{n-2}v_{(a,1)} + (n\!-\!a\!-\!2)\lambda v_{(a,0)}}{\frac{(a-1)\lambda}{2(n-1)} + \frac{\lambda}{2} + \frac{(a-1)(n-a-2)\lambda}{n-2} + (n\!-\!a\!-\!2)\lambda} \!\!\!\\
    v_{(a,0)} = &\frac{\lambda_e + \frac{a(n-a-1)\lambda}{n-2}v_{(a+1,0)} + a\lambda v_{(a-1,1)}}{\frac{a\lambda}{2(n-1)} + \frac{a(n-a-1)\lambda}{n-2} + a\lambda}. \label{second set}
\end{align}

We can see that $v_{(a-1,1)} < v_{(a,0)}$, due to the higher update rate for the position in $\mathcal{N}_{s}$. Hence, at each stage, we can write $v_{(a,0)}$ in terms of $v_{(a,1)}$ and $v_{(a+1,0)}$. Then, we can obtain an upper bound for the recursive equation by replacing $v_{(a,1)}$ with $v_{(a+1,0)}$. In order to simplify notation, we define
\begin{align}
    c^{(a)} =& \frac{(a-1)\lambda}{2(n-1)} + \frac{\lambda}{2} + \frac{(a-1)(n-a-2)\lambda}{n-2} + (n-a-2)\lambda, \\
    d^{(a)} =& \frac{a\lambda}{2(n-1)} + \frac{a(n-a-1)\lambda}{n-2} + a\lambda.
\end{align}
Then, we substitute the equation for $v_{(a-1,1)}$ in \eqref{second set} as
\begin{align}
    \!\!\!\!v_{(a,0)} = &\left(1-\frac{(n-a-2)a \lambda^2}{c^{(a)}d^{(a)}}\right)^{-1}\Biggl(\frac{\lambda_e}{d^{(a)}}\left[1+\frac{a\lambda}{c^{(a)}}\right] \Biggr.\notag\\
    &\Biggl.+\frac{\frac{a(n-a-1)\lambda}{n-2}v_{(a+1,0)}}{d^{(a)}} + \frac{\frac{a(a-1)(n-a-2)\lambda^2}{n-2}v_{(a,1)}}{c^{(a)}d^{(a)}}\Biggr)\!
\end{align}
Now, we replace $v_{(a,1)}$ with $v_{(a+1,0)}$ and write an upper bound and simplify to get,
\begin{align}
    v_{(a,0)} \leq \frac{1}{z^{(a)}}(\lambda_e x^{(a)}+y^{(a)}v_{(a+1,0)}),
\end{align}
where $x^{(a)} = c^{(a)}+a\lambda$, $y^{(a)} = \frac{a(n-a-1)\lambda c^{(a)}}{n-2} + \frac{a(a-1)(n-a-2)\lambda^2}{n-2}$ and $z^{(a)} = c^{(a)}d^{(a)}-(n-a-2)a \lambda^2$. Then, we write out the recursive upper bound and obtain,
\begin{align}\label{eq: application recursion}
    v_{(1,0)} \leq \lambda_e \sum_{i=1}^{n-2}\frac{x^{(i)}}{z^{(i)}}\prod_{j=1}^{i-1} \frac{y^{(j)}}{z^{(j)}} + \prod_{j=1}^{n-2} \frac{y^{(i)}}{z^{(i)}}v_{(n-1,0)}.
\end{align}
where $v_{(n-1,0)}$ is a constant, and depends on $v_{\mathcal{N}}$ and $v_{(n-2,1)}$. Using the recursive equations, we can obtain $v_{(n-1,0)}$ and see that it is a constant.
Now, we want to simplify \eqref{eq: application recursion} in order to find a simple upper bound. First, we simplify $\frac{y^{(i)}}{z^{(i)}}$. In order to simplify notation, we define, $n_1 = \frac{i(i-1)(n-i-1)}{2(n-1)(n-2)}$, $n_2 = \frac{i(n-i-1)(2n-2i-1)}{2(n-2)}$, $n_3 = \frac{(i-1)i(n-i-2)(n-i-1)}{(n-2)^2}$, $d_1 = \frac{i(4n-7)}{4(n-1)}$, $d_2 = \frac{i(n-i-2)(2n-3)}{2(n-2)}$, $d_3 = \frac{(i-1)i(2n-2i-3)}{2(n-1)(n-2)}$ and $d_4 = \frac{(i-1)i(n-i-2)(n-i-1)}{(n-2)^2}$.
Then, for $i < n-2$, we have,
\begin{align}
    \frac{y^{(i)}}{z^{(i)}} &\approx \frac{n_1+n_2+n_3}{d_1+d_2+d_3+d_4}\\
    &\leq \frac{n_1+n_2+n_3}{d_2+d_4}\\
    &\approx \frac{\frac{(n-i-1)(2n-2i-1)}{2(n-2)}+\frac{(i-1)(n-i-2)(n-i-1)}{(n-2)^2}}{\frac{(n-i-2)(2n-3)}{2(n-2)}+\frac{(i-1)(n-i-2)(n-i-1)}{(n-2)^2}}\\
    &= \frac{1+\frac{(2n-2i-1)(n-2)}{2(i-1)(n-i-2)}}{1+\frac{(2n-3)(n-2)}{2(i-1)(n-i-1)}}\\
    &\approx \frac{1+\frac{n-2}{i-1}}{1+\frac{n-2}{i-1} + \frac{n-2}{n-i-1}}\\
    &\leq 1.
\end{align}
Also, $\frac{y^{(n-2)}}{z^{(n-2)}} \leq 1$, thus,  $\prod_{j=1}^{i-1} \frac{y^{(j)}}{z^{(j)}} \leq 1, \forall i$, and \eqref{eq: application recursion} becomes,
\begin{align}
    v_{(1,0)} \leq \lambda_e \sum_{i=1}^{n-2}\frac{x^{(i)}}{z^{(i)}} + \prod_{j=1}^{n-2} \frac{y^{(i)}}{z^{(i)}}v_{(n-1,0)}.
\end{align}

We now simplify $\frac{x^{(i)}}{z^{(i)}}, i < n-2$, as
\begin{align}
    \frac{x^{(i)}}{z^{(i)}} &= \frac{n-2 + \frac{i-1}{2(n-1)}+\frac{1}{2}+\frac{(i-1)(n-i-2)}{n-2}}{d_1+d_2+d_3+d_4}\\
    &\leq \frac{n-2 + 1 + \frac{(i-1)(n-i-2)}{n-2}}{d_2+d_3+d_4}\\
    &= \frac{1}{\lambda}\frac{1 + \frac{1}{n-2} + \frac{(i-1)(n-i-2)}{(n-2)^2}}{\frac{i(2n-3)(n-i-2)}{2(n-2)^2}+\frac{(i-1)i(n-i-1)(n-i-2)}{(n-2)^3}}\label{eq: xi/zi first}\\
    &\leq \frac{1}{\lambda}\frac{\frac{5}{4}}{\frac{i(n-i-2)}{n-2}+\frac{(i-1)i(n-i-1)(n-i-2)}{(n-2)^3}}\label{eq: xi/zi second}\\
    &\leq \frac{1}{\lambda}\frac{\frac{5}{4}}{\frac{i(n-i-2)}{n-2}}\label{eq: xi/zi third}\\
    &= \frac{5}{4}\frac{1}{\lambda}\left(\frac{1}{i}+\frac{1}{n-i-2}\right).
\end{align}
We go from \eqref{eq: xi/zi first} to \eqref{eq: xi/zi second} by $\frac{1}{n-2} \approx 0$ and $\frac{(i-1)(n-i-2)}{(n-2)^2} \leq \frac{1}{4}$ when $i < n-2$. Further, $\frac{x^{(n-2)}}{z^{(n-2)}}\leq \frac{1}{\lambda} \leq \frac{5}{4\lambda}$.
Hence, we have
\begin{align}
    v_{(1,0)} &\leq \frac{5}{4}\frac{\lambda_e}{\lambda} \sum_{i=1}^{n-3}\left(\frac{1}{i}+\frac{1}{n-i-2}\right) + C\\
    &\leq \frac{5}{2}\frac{\lambda_e}{\lambda}\log{n} + C',\label{eq: application result}
\end{align}
where $C' = v_{(n-1,0)} + \frac{5}{2}\gamma + \frac{5\lambda_e}{4\lambda}$ where $\gamma \approx 0.577$ is the Euler-Mascheroni constant. Hence, $v_{(1,0)} = O(\log{n})$. Since all but one node have the same version age, the version age of a single node in such a network scales as $O(\log{n})$.

\begin{figure}
    \centering
    \includegraphics[width=0.65\linewidth]{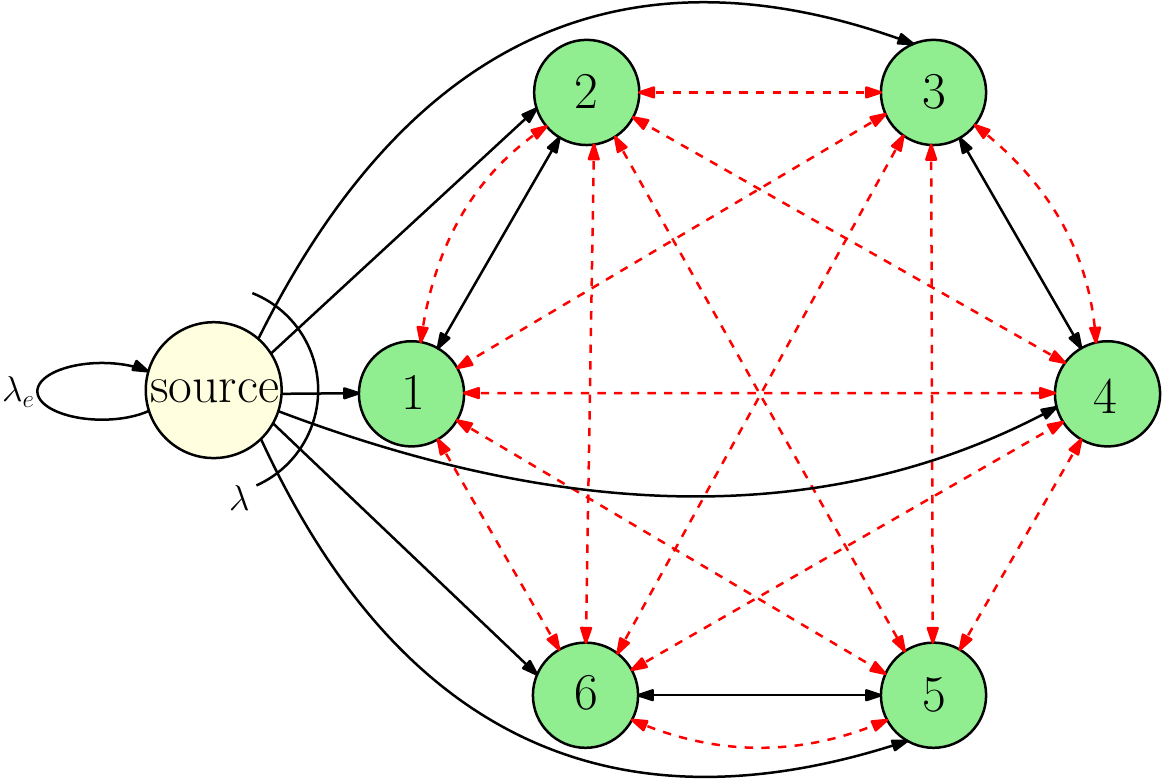}
    \caption{A disconnected gossiping network which has $6$ nodes, and nodes in two adjacent positions gossip with each other, but not with nodes in any other position, i.e., a node in each position can gossip with a node only in one other position. This is represented by solid black double-sided arrows. Further, a node in any position can exchange positions with nodes in any other position in the network. This is represented by red dashed arrows. $\{1,2\}$, $\{3,4\}$ and $\{5,6\}$ are adjacent node pairs. $\{1\}$ is a single node of the node pair $\{1,2\}$.}
    \label{fig: disconnected network with full mobility}
    \vspace*{-0.5cm}
\end{figure}

\section{Disconnected Network with Full Mobility}\label{sec: disconnected network}
In this section, we consider a gossiping network where a node in any position can communicate with only one node in an adjacent position, i.e., a $1$-regular graph, as shown in Fig.~\ref{fig: disconnected network with full mobility}. A node in a particular position in the network can also exchange positions with any other node in the network. Therefore, $M_i = \mathcal{N} \backslash i, \forall i \in \mathcal{N}$. The rate of all node exchanges is the same and is set as $\lambda_m$. Hence, $\lambda_{ij}^m = \lambda_{ji}^m = \lambda_m, i,j \in \mathcal{N}, i \neq j$. Any set of positions in this network can be uniquely defined by the number of adjacent node pairs and single nodes of an adjacent pair it contains, as described in Fig.~\ref{fig: disconnected network with full mobility}. Hence, each set of positions with the same number of adjacent node pairs and single nodes in it have the same version age. Thus, each set of positions can be represented as $(\alpha,\beta)$ where $\alpha$ is the number of adjacent node pairs and $\beta$ is the number of single nodes. In order to find an upper bound for the version age of a single node in the network, we will obtain a recursive upper bound on the set $(a,0)$ in terms of the set $(a+1,0)$ using the fact that given $2a$ nodes, the set with the highest version age is the set with $a$ node pairs, and given $2a+1$ nodes, the set with the highest version age is $(a,1)$. These statements are true due to the fact that if we consider sets of nodes of the same size, then the set of positions with the highest version age is the one which has the least amount of gossip coming into it from nodes outside the set. The impact of mobility for any two sets of the same size is the same due to the full mobility of the network and the fact that due to the symmetry of the network, each node in the network has the same version age. We then start the recursion by writing the version age of a single node in terms of a node pair,
\begin{align}
    v_{(0,1)} = \frac{\lambda_e+\lambda v_{(1,0)}}{\frac{\lambda}{n}+\lambda},
\end{align}
and then write $(a,0)$ in terms of $(a+1,0)$ to find a recursive upper bound. We write $(a,0)$ in terms of $(a+1,0)$ for $a=1$ to $a = \frac{n}{2}-2$ as follows,
\begin{align}
    v_{(a,0)} &= \frac{\lambda_e+2a(n-2a)\lambda_m v_{(a-1,2)}}{\frac{2a\lambda}{n}+2a(n-2a)\lambda_m},\\
    v_{(a-1,2)} &\leq \frac{\lambda_e+2\lambda v_{(a,1)} + (2\lambda_m +h^{(a)})\lambda_m) v_{(a-1,2)}}{\frac{2a\lambda}{n}+2\lambda+2\lambda_m + h^{(a)}},\\
    v_{(a,1)} &\leq \frac{\lambda_e + \lambda v_{(a+1,0)} + g^{(a)} v_{(a,1)}}{\frac{(2a+1)\lambda}{n}+\lambda+g^{(a)}},
\end{align}
where $h^{(a)} = 2(a-1)(n-2(a+1))\lambda_m$ and $g^{(a)} = 2a(n-2(a+1))\lambda_m$.
Now, to simplify notation, we define, $b^{(a)} = \frac{2a\lambda}{n}+2a(n-2a)\lambda_m$, $c^{(a)} = \frac{(2a+1)\lambda}{n}+\lambda+2a(n-2(a+1))\lambda_m$, $d^{(a)} = \frac{2a\lambda}{n}+2\lambda+2\lambda_m+2(a-1)(n-2(a+1))\lambda_m$, $f^{(a)} = 1-\frac{2a(n-2(a+1))\lambda_m}{c^{(a)}}$ and $g^{(a)} = 1-\frac{2(a-1)(n-2(a+1))\lambda_m}{d^{(a)}}$. Then,
\begin{align}
    v_{(a,0)} \leq \lambda_e x^{(a)} + y^{(a)}v_{(a+1,0)},
\end{align}
where
\begin{align}
    x^{(a)} &= \frac{1}{b^{(a)}}+\frac{2a(n-2a)\lambda_m}{b^{(a)}d^{(a)}g^{(a)}}+\frac{4a(n-2a)\lambda_m\lambda}{b^{(a)}c^{(a)}d^{(a)}f^{(a)}g^{(a)}},\\
    y^{(a)} &= \frac{4a(n-2a)\lambda_m\lambda^2}{b^{(a)}c^{(a)}d^{(a)}f^{(a)}g^{(a)}}.
\end{align}
Suppose $\lambda_m = \frac{\lambda}{f(n)}$. Then, we get,
\begin{align}
    x^{(a)} &\leq \frac{1}{\frac{2a\lambda}{n}+2a(n-2a)\frac{\lambda}{f(n)}}\left(1+\frac{\frac{3a(n-2a)}{f(n)}}{1+\frac{a}{n}+\frac{1}{f(n)}}\right),\label{eq: x^{(a)}}\\
    y^{(a)} &= \frac{1}{(1+\frac{2f(n)}{n(n-2a)})(1+\frac{2a+1}{n})(1+\frac{1}{f(n)}+\frac{a}{n})}.\label{eq: y^{(a)}}
\end{align}
Hence, we can say that,
\begin{align}
    v_{(1,0)} \leq \lambda_e \sum_{i=1}^{\frac{n}{2}-2}x^{(i)}\prod_{j=1}^{i-1}y^{(j)} + v_{(\frac{n}{2}-1,0)}\prod_{j=1}^{\frac{n}{2}-1}y^{(j)}.
\end{align}
Define $\alpha_i = \prod_{j=1}^{i-1}y^{(j)}$. We note that $v_{(\frac{n}{2}-1,0)}$ is a constant. Then, the version age of a single node is upper bounded as,
\begin{align}
    v_{(0,1)} &\leq \frac{\lambda_e}{\lambda}+\lambda_e \sum_{i=1}^{\frac{n}{2}-2}x^{(i)}\prod_{j=1}^{i-1}y^{(j)} + v_{(\frac{n}{2}-1,0)}\prod_{j=1}^{\frac{n}{2}-1}y^{(j)}\\
    &\leq \frac{\lambda_e}{\lambda}+\frac{\lambda_e}{\lambda} \sum_{i=1}^{\frac{n}{2}-2}x^{(i)}\alpha_i + C,\label{eq: disconnected final sum}
\end{align}
where C = $v_{(\frac{n}{2}-1,0)}$, since $y^{(j)}\leq 1, \forall j$. We now find the upper bound scaling for two regions for $f(n)$. For each region, we bound $x^{(a)}$ and $y^{(a)}$ appropriately and then find the sum in \eqref{eq: disconnected final sum}. The first region is  $f(n) = k \in \mathbb{R}$. Starting with \eqref{eq: x^{(a)}}, 
\begin{align}
    x^{(a)} \leq& \frac{\frac{3a(n-2a)}{f(n)}}{\frac{2a\lambda}{n}+2a(n-2a)\frac{\lambda}{f(n)}} \leq \frac{3}{2\lambda},\label{eq: x^{(a)} constant}
\end{align}
and starting from \eqref{eq: y^{(a)}}, $y^{(a)}$ can be upper bounded as,
\begin{align}
    y^{(a)} \leq& \frac{1}{1+\frac{1}{k}+\frac{a}{n}}\leq \frac{k}{k+1}.
\end{align}
We substitute this back in \eqref{eq: disconnected final sum},
\begin{align}
    v_{(0,1)} \leq& \frac{\lambda_e}{\lambda}+\frac{3\lambda_e}{2\lambda} \sum_{i=1}^{\frac{n}{2}-2}\prod_{j=1}^{i-1}\frac{k}{k+1} + C\\
    =& \frac{\lambda_e}{\lambda}+\frac{3\lambda_e}{2\lambda} \sum_{i=1}^{\frac{n}{2}-2}\bigl( \frac{k}{k+1} \bigr)^{i-1} + C\\
    \leq& \frac{\lambda_e}{\lambda}+\frac{3\lambda_e}{2\lambda}k + C',\label{eq: first region}
\end{align}
where $C' = C + \frac{3\lambda_e}{2\lambda}$. Hence, the version age is upper bounded by a constant in this case. In the second region, we choose $f(n)$ such that $f(n) = O(n)$ and $1 = o(f(n))$. First, we note that we use the same bound for $x^{(a)}$ as in \eqref{eq: x^{(a)} constant}. Starting at \eqref{eq: y^{(a)}}, we bound $y^{(a)}$ as follows,
\begin{align}
    y^{(a)} \leq \frac{1}{1+\frac{1}{f(n)}}\frac{1}{1+\frac{2a}{n}}.
\end{align}
Then, we find $\alpha_i$ by taking negative logarithm on both sides,
\begin{align}
    -\log{\alpha_i} =& \sum_{j=1}^{i-1} \log\left(1+\frac{1}{f(n)}\right)+\log\left(1+\frac{2j}{n}\right)\\
    =& \sum_{j=1}^{i-1} \frac{1}{f(n)}+\frac{2j}{n}\\
    \approx& \frac{i}{f(n)} + \frac{i^2}{n}.
\end{align}
Hence, we get,
\begin{align}
    \alpha_i \approx e^{-(\frac{i}{f(n)} + \frac{i^2}{n})}.\label{eq: alpha main}
\end{align}
We now further upper bound $\alpha_i$ in two different ways. First, 
\begin{align}
    \alpha_i \leq e^{-\frac{i}{f(n)}}.
\end{align}
Substituting this back in \eqref{eq: disconnected final sum}, we get,
\begin{align}
    v_{(0,1)} \leq& \frac{\lambda_e}{\lambda}+\frac{3\lambda_e}{2\lambda} \sum_{i=1}^{\frac{n}{2}-2}e^{-\frac{i}{f(n)}} + C\\
    =& \frac{\lambda_e}{\lambda}+\frac{3\lambda_e}{2\lambda} f(n) + C.\label{eq: second region bound 1}
\end{align}
using Riemann sums. Next, we bound \eqref{eq: alpha main} in the second way,
\begin{align}
    \alpha_i \leq e^{-\frac{i^2}{n}}.
\end{align}
Substituting this back in \eqref{eq: disconnected final sum}, we get,
\begin{align}
    v_{(0,1)} \leq& \frac{\lambda_e}{\lambda}+\frac{3\lambda_e}{2\lambda} \sum_{i=1}^{\frac{n}{2}-2}e^{-\frac{i^2}{n}} + C\\
    \leq& \frac{\lambda_e}{\lambda}+\frac{3\lambda_e}{2\lambda} \frac{\sqrt{\pi}}{2}\sqrt{n} + C.\label{eq: second region bound 2}
\end{align}
using the same method as in \cite{buyukates22ClusterGossip}. Hence, we combine \eqref{eq: second region bound 1} and \eqref{eq: second region bound 2} to get the following upper bound for the version age of a single node for the second region of $f(n)$,
\begin{align}
    v_{(0,1)} = O(\min(f(n),\sqrt{n})).\label{eq: second region}
\end{align}
We now combine the results in \eqref{eq: first region} and \eqref{eq: second region}, and obtain
\begin{align}
    v_{(0,1)} = O(\min(f(n),\sqrt{n})).\label{eq: disconnected network result}
\end{align}
Hence, if $f(n) = \Omega(\sqrt{n})$, then we see that the version age of a single node scales as $O(\sqrt{n})$. If $f(n) = o(\sqrt{n})$, then the version age scaling is $O(f(n))$.

\section{Discussion}

\begin{remark}
    In a fully connected network, adding mobility does not improve the version age of nodes in the network. To prove this, we can show that, even if every node can exchange its position with every other node in the network, the version age of a single node scales as $\log{n}$. In this case, due to the symmetry in the network, any two sets formed in the recursion will have equal version ages if they are of the same size. Hence, if $|S| = k$, then in \eqref{eq: recursive equation}, we can replace $v_S$ and $v_{S \cup \{j\} \backslash \{i\}}$ with $v_k$ and $v_{S \cup \{j\}}$ with $v_{k+1}$. Then, we see that the extra mobility term gets cancelled, and we are left with the recursion from \cite{yates21gossip}. From this, we conclude that, if all nodes can exchange positions with any other node in the network, then the version age scaling is still $\log{n}$. Hence, the version age of a fully connected network does not improve with mobility.
\end{remark}

\begin{remark}
    Nodes being able to exchange positions does not improve the version age of every position in the network, and therefore, does not improve the version age of every node in the network. This is because, nodes which are in advantageous positions in networks that have no mobility, may not be in the same position for the entire time horizon in networks with mobility, but may travel to positions which typically see less updates from the rest of the network or from the source.
\end{remark}

\begin{remark}
    The result in \eqref{eq: disconnected network result} tells us that if $\lambda_m = \Omega{(\frac{\lambda}{\log{n}})}$, then the version age scaling is the same as a fully connected network. Moreover, full mobility improves the version age of a node in the disconnected network described in Section~\ref{sec: disconnected network} to be at least as good as the version age in a ring network.
\end{remark}

\section{Numerical Simulations}\label{sec: numerical simulations}
\begin{figure}
    \centering
    \includegraphics[scale = 0.4]{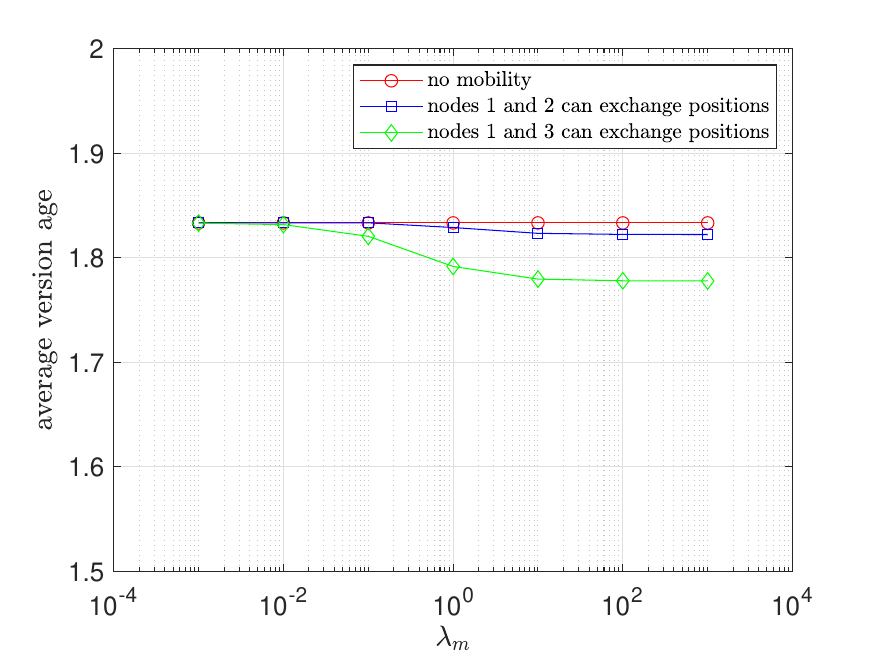}
    \caption{A comparison of average version age in the toy example. We compare the cases when there is no mobility, represented by red, where nodes $1$ and $2$ can exchange positions, represented by blue, and where nodes $1$ and $3$ can exchange positions, represented by green.}
    \label{fig: toy example 1}
    \vspace*{-0.2cm}
\end{figure}

We first plot the results of the toy example for different values of $\lambda_m$ while choosing $\lambda_e = \lambda = 1$. We choose $\lambda_m$ among $[0.001,0.01,0.1,1,10,100,1000]$. In Fig.~\ref{fig: toy example 1}, we plot the average version age for each of the three networks. Since the version age of positions in the case of no mobility will not depend on $\lambda_m$, the average age is constant. In both cases of node exchanges, the average version age decreases due to mobility. We can clearly see that as $\lambda_m$ increases, the average version age decreases. Moreover, the average version ages are quite similar for small values of $\lambda_m$. However, when nodes in positions $1$ and $3$ exchange positions with each other, the version age decreases significantly more than when nodes in positions $1$ and $2$ exchange positions. 

\begin{figure}
    \centering
    \includegraphics[scale = 0.4]{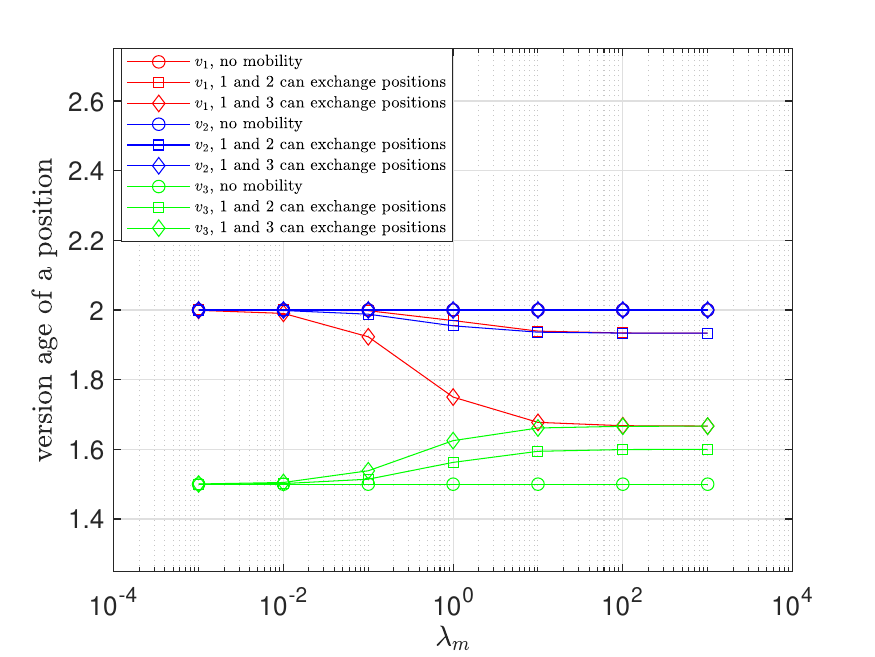}
    \caption{A comparison of version age of individual nodes in the toy example. The first position is given in red, the second position is given in blue and the third position is given in green.}
    \label{fig: toy example 2}
    \vspace*{-0.4cm}
\end{figure}

In Fig.~\ref{fig: toy example 2}, we plot the individual version age of each position in all three networks. We see that in the case where nodes $1$ and $3$ exchange positions, the version age of position $1$ decreases significantly whereas the version age of position $3$ increases significantly and the version age of position $2$ remains the same. Moreover, as $\lambda_m$ becomes large, the version age of positions $1$ and $3$ become equal. This is because, on average, both nodes are in each position half the time. When nodes in positions $1$ and $2$ exchange, then the version ages of positions $1$ and $2$ decrease less in magnitude when compared to the previous case. Also, the average version age of position $3$ increases less than the previous case. This is because the node in position $2$ gets less updates, and  it moves to position $1$ having high version age for some time. Hence, the version updates of position $3$ become less frequent.

Next, we simulate the network considered in Section~\ref{sec app}. We choose $\lambda_e = \lambda = \lambda_m = 1$. We see that having mobility in the network improves the version age of the nodes and the upper bound we obtain is close to the simulated version age.

\begin{figure}
    \centering
    \includegraphics[scale = 0.4]{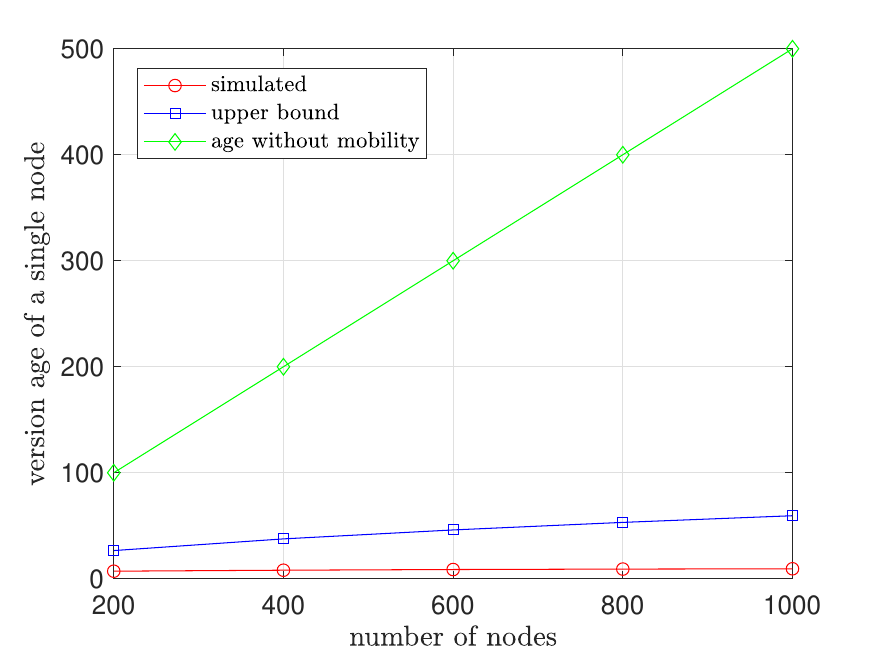}
    \caption{A comparison of the mobility gossip network considered in Section~\ref{sec: disconnected network}. We compare the average version age of a single node when there is mobility, when there is no mobility and the upper bound we found in \eqref{eq: disconnected network result}.}
    \label{fig: disconnected plot}
    \vspace*{-0.4cm}
\end{figure}

Finally, we simulate the network considered in Section~\ref{sec: disconnected network} and plot the results in Fig.~\ref{fig: disconnected plot}. We choose $\lambda_e = \lambda = 1$ and $\lambda_m = \frac{1}{n}$. We see that the simulated version age of the network follows logarithmic scaling. Our upper bound follows square root scaling and the version age scales linearly when there is no mobility. From these simulations, we see that the version age scaling of a single node in the network considered in Section~\ref{sec: disconnected network} scales as $O(\log{n})$ if $f(n) = O(n)$.

\begin{figure}[t]
    \centering
    \includegraphics[scale = 0.4]{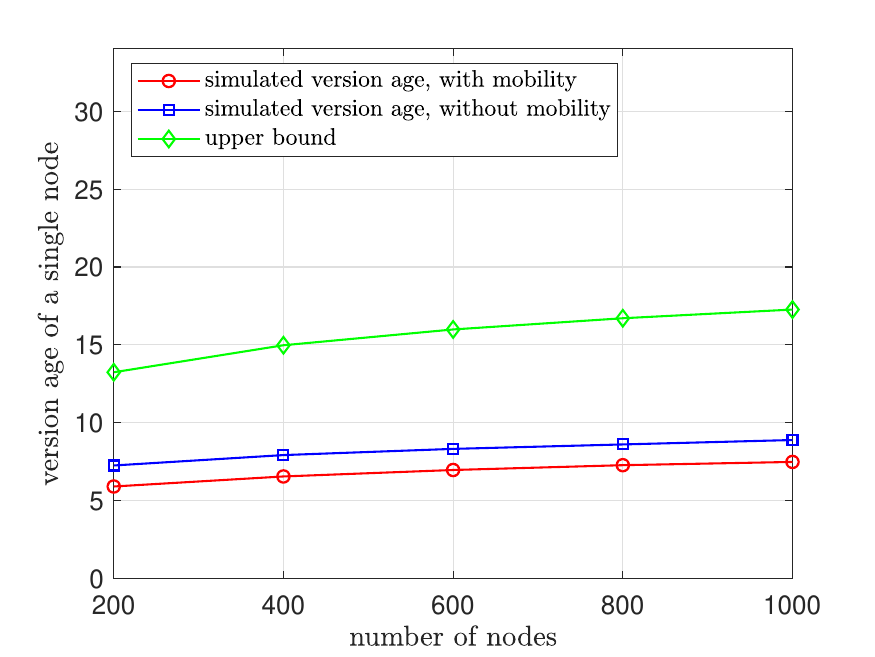}
    \caption{A comparison of the mobility gossip network considered in Sec.~\ref{sec app}. We compare the average version age of a single node when there is mobility, when there is no mobility and the upper bound we found in \eqref{eq: application result}.}
    \label{fig: application plot}
    \vspace*{-0.2cm}
\end{figure}

\bibliographystyle{unsrt}
\bibliography{refs}

\end{document}